\documentclass[11pt,aps,superscriptaddress,floatfix]{revtex4}
\usepackage{graphicx}
\usepackage{epsfig}
\usepackage{amsmath,amssymb}
\usepackage{color}
\newcommand{\be}{\begin{eqnarray}}
\newcommand{\ee}{\end{eqnarray}}
\newcommand\del{\partial}
\def\conj#1{{{#1}^{*}}}

\newcommand{\mat}{\left ( \begin{array}{cc}}
\newcommand{\emat}{\end{array} \right )}
\newcommand{\matf}{\left ( \begin{array}{cccc}}
\newcommand{\ematf}{\end{array} \right )}            
\newcommand{\matt}{\left \begin{array}{ccc}}
\newcommand{\ematt}{\end{array} \right )}
\newcommand{\vect}{\left ( \begin{array}{c}}
\newcommand{\evect}{\end{array} \right )}

\newcommand{\nn}{\nonumber } 

\newcommand{\hz}{\hat{z}}
\newcommand{\hzs}{\conj{\hat{z}}}

\newcommand{\hmu}{\hat{\mu}}
\def\conj#1{{{#1}^{*}}}

\definecolor{Bittersweet}   {cmyk}{0,0.75,1,0.24}

\begin{document}

\title{Phase Diagram of the Dirac Spectrum at Nonzero Chemical Potential}

\author{J.C. Osborn}
\affiliation{Argonne Leadership Computing Facility, 9700 S. Cass Avenue,
Argonne, IL 60439, USA}
\affiliation{Center for Computational Science, Boston University,
Boston, MA 02215, USA}

\author{K. Splittorff}
\affiliation{The Niels Bohr Institute, Blegdamsvej 17, DK-2100, Copenhagen {\O}, Denmark}

\author{J.J.M. Verbaarschot}
\affiliation{Department of Physics and Astronomy, SUNY, Stony Brook,
 New York 11794, USA}

\date{\today}
\begin{abstract}
  The Dirac spectrum of QCD with dynamical fermions 
 at nonzero chemical potential is characterized
  by three regions, a region with a constant eigenvalue density, a region
  where the eigenvalue density shows oscillations that grow exponentially
  with the volume and the remainder of the complex plane where the 
  eigenvalue density is zero. In this paper we derive the phase diagram of
 the Dirac spectrum from a chiral Lagrangian. We show that the constant
 eigenvalue density corresponds to a pion condensed phase while the
strongly oscillating region is given by a kaon condensed phase. The normal
phase with nonzero chiral condensate but vanishing Bose condensates
coincides with the region of the complex plane where there are no eigenvalues.
\end{abstract}
\maketitle

%}
%\tableofcontents
%\widetext
%\onecolumn

\newpage

%%%%%%%%%%%%%%%%%%%%%%%%%%%%%%%%%%%%%%%%%%%%%%%%%%%%%%%%%%%%%%%%%%%%%%%%%%%%%%
\section{Introduction}

Because of the Banks-Casher relation \cite{BC} between the chiral condensate
and the spectral density of the QCD Dirac operator, lattice QCD studies
have had a long lasting interest in the Dirac spectrum. The eigenvalue
correlations have been well understood for 
an (anti-)Hermitian Dirac operator. In the
chirally broken phase, the eigenvalue density around zero is nonzero
in the thermodynamic limit resulting in a discontinuity of the chiral
condensate. In the chirally restored phase, a gap develops around zero
so that the condensate remains a smooth function of the mass in the
thermodynamic limit and therefore vanishes in the massless limit.

At nonzero quark chemical potential, $\mu$, the Dirac spectrum  
also received considerable attention in the lattice QCD literature 
\cite{all,gibbs,Baillie,Vink,Davies,LKS,Schafer,Markum,Bittner}.
In quenched lattice simulations it was found that
the Dirac eigenvalues are distributed more or less homogeneously inside
a strip that is symmetric about the imaginary axis. 
In particular, this implies
that the quenched chiral condensate vanishes in the massless
limit for arbitrarily small chemical potential \cite{all}.
On the other hand, because at very low temperatures 
the free energy of QCD with dynamical quarks 
does not depend on the quark chemical 
potential below a third of the nucleon mass,
chiral symmetry 
 must remain broken at small nonzero chemical potential if it is broken
at zero chemical potential. Naively, the fermion 
determinant will make matters worse (an even smoother behavior of the
condensate at $m=0$) because eigenvalues will be repelled
from the position of the mass \cite{all}, 
but as will be discussed below, it is precisely the
phase of the fermion determinant that will lead to a nonzero 
chiral condensate. 

A partial resolution of this paradox appeared \cite{gibbs} 
soon after the publication of \cite{all}. Based on a one-dimensional 
$U(1)$ lattice model it was concluded that phase fluctuations
of the quark determinant are essential 
to obtain a nonzero chiral condensate. In the same paper, it was found that
the width of the strip of eigenvalues is equal to the quark mass when
$\mu = m_\pi/2$. This has been verified by lattice simulations
\cite{Davies,LKS}. Recently, this result was derived from a 
chiral Lagrangian for the Dirac spectrum of quenched QCD \cite{TV}. 
However, the paradox
of how a $\mu$-dependent spectral density can lead to both a
$\mu$-independent partition function and a 
discontinuity of the chiral condensate remained unsolved and was later coined
\cite{cohen} as the ``Silver Blaze Problem.''

The resolution of the ``Silver Blaze Problem'' became possible because of two 
further developments. First, the idea to explain the failure of the
quenched approximation at $\mu \ne 0$ in a random matrix model with
the symmetries of
QCD at $\mu =0$ \cite{misha}. 
Second, the introduction of a mathematically simpler
version of this model that made it possible to derive exact analytical
expressions for the spectral density of the QCD Dirac operator with
dynamical quarks \cite{O,AOSV}. Using these results it was shown in
\cite{OSV,OSVmicro} that a strongly oscillating contribution to 
the spectral density is responsible for the discontinuity in the chiral
condensate.

The random matrix results are valid for QCD in the microscopic
 limit \cite{SV,V}  where the combinations
\be
\hat m \equiv  m \Sigma V \qquad {\rm and} \qquad \hat \mu^2 \equiv \mu^2 F^2 V
\label{micro}
\ee
are kept fixed in the thermodynamic limit 
(Here, $m$ denotes any of the quark masses or the mass
scale for which the Dirac spectrum is calculated, $\mu$ is the chemical
potential, $\Sigma$ is the magnitude of the chiral condensate and $F$ is
 the pion decay constant).  Since in this limit, the contribution 
of the zero momentum modes factorizes from the partition function \cite{Vplb}, 
it is also equivalent to the $\epsilon$ regime of chiral perturbation theory
\cite{GL}. 
Random matrix predictions for the quenched spectral density
\cite{SplitVerb2,Ake} have been compared successfully to lattice gauge
simulations for the microscopic domain \cite{AW,OW,BW}.

In this paper we show that the microscopic results for the
 unquenched Dirac spectrum are characterized by three
 different regions or ``phases'': a phase where there are no eigenvalues, 
 a phase where the eigenvalue density is equal to the quenched result, 
 and a phase where the Dirac spectrum is strongly oscillating.
The aim of the present paper is to derive the phases of the Dirac spectrum
 from a mean field analysis of the chiral Lagrangian for the Dirac spectrum of
 QCD at $\mu \ne 0$.

In the mean field limit one evaluates the extrema of the static part of the
chiral Lagrangian so that the leading order
contribution in the $p$-expansion of chiral perturbation theory
is exactly given by
the large $\hat m$ and large $\hat \mu$ limit of the microscopic limit of
QCD. The classification of the spectral density of the Dirac
operator is therefore valid beyond the microscopic regime up to energies
that are well below the QCD scale.
Alternatively, the locus of these phases in the 
complex eigenvalue plane can  be obtained \cite{OSV} from 
the asymptotic properties of the
complex Laguerre polynomials used to solve the corresponding random matrix
model.

In section \ref{sec:quench} we review the calculation of the phase diagram
 of the quenched Dirac spectrum.
The phase diagram of the Dirac spectrum for QCD with one
dynamical flavor is discussed in section \ref{sec:unquench},
 and the dependence of the free
energy on the number of replicas is discussed in section \ref{sec:nrep}.
Concluding remarks are made in
section \ref{sec:nrep}. A brief discussion of some of the 
 results of this paper appeared in the proceedings
\cite{osvchina}.

\section {Phase Diagram for the Quenched Dirac Spectrum}
\label{sec:quench}

In this section we review the mean field analysis of the phases
of the quenched Dirac spectrum originally given in \cite{TV} using the 
supersymmetric method. In this paper  we evaluate the quenched spectral 
density  by means of the replica trick  given by \cite{girko,misha}
\be
\rho^{N_f=0}(z,z^*;\mu) &=& \lim_{n\to 0} 
\frac 1V \frac{1}{n} \del_z \del_{z^*} \log 
Z^n(z,z^*;\mu),\nn\\
Z^n(z,z^*;\mu) &=&\langle {\det}^n(D +\gamma_0\mu +z){\det}^n(D -\gamma_0\mu +z^*)\rangle ,
\ee 
where $\langle \ldots \rangle$ denotes the quenched ensemble average.
In the mean field limit the result for $\rho^{N_f=0}(z,z^*;\mu)$ before
taking the replica limit, $n\to 0$, 
does not depend on $n$ and the limit $n\to 0$ can be taken trivially. 
Therefore it is sufficient
to analyze the partition function for $n=1$ which, as is illustrated by the
above rewriting of the partition function, is equivalent to QCD at
nonzero isospin chemical potential with complex quark masses. 
The replica argument can be made rigorous in the microscopic limit
\cite{KanzieperPRL,SplitVerb1} through the Toda lattice equation which links
the generating functionals, $Z^n$ with different replica index $n$. Using the
Toda lattice equation one can derive an exact identity \cite{SplitVerb2,AOSV}
for the quenched microscopic spectral density in the sector with topological
charge $\nu$
\be
\rho^{N_f=0}(\hz,\hzs;\hmu) =\frac 12 \hz\hzs
Z^{n=1}(\hz,\hzs;\hmu)Z^{n=-1}(\hz,\hzs;\hmu). 
\ee
Since the microscopic limit of the partition function with a pair of
conjugate bosonic quarks, $Z^{n=-1}(z,z^*;\mu)$, 
 given by \cite{SplitVerb2,AOSV}
\be
Z^{n=-1}(\hz,\hzs;\hmu)  = |\hz|^{2\nu} 
\frac{1}{\hmu^2} 
\exp\left(- \frac{(\hat z^2 + \hat z^{*\,2})}{8 \hat \mu^2 }  \right)
K_\nu \left( \frac{|\hat z|^2}{4 \hat \mu^2}  \right),
\label{zmin1}
\ee
is a smooth function of $z$, the phase structure of the spectral
density is given by  $Z^{n=1}(z,z^*;\mu)$ (in agreement with the
conclusion from the mean field replica trick). 

 \begin{center}
\begin{figure}[!t]
  \unitlength1.0cm
  \begin{center}
  \centerline{\epsfig{file=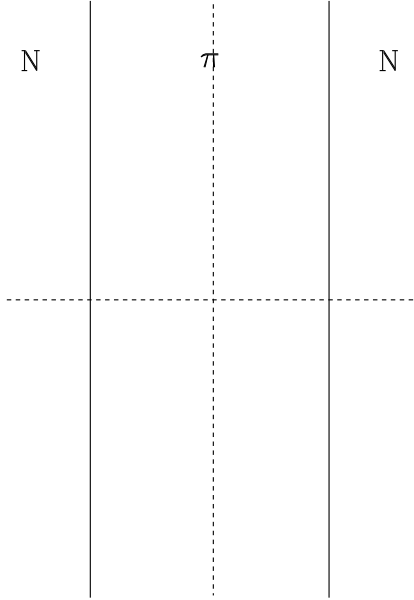,height=8cm}}
  \end{center}
  \caption{ 
Phase diagram of the quenched Dirac spectrum in the complex $z$ plane.
The support of
the quenched spectrum is between the two solid vertical black lines. 
The labels N and $\pi$ refer to
the normal and  pion condensed phase, respectively.}
 \label{fig:rhoq}
\end{figure}
\end{center}

We will examine the partition function $Z^{n=1}(z,z^*;\mu)$ to leading 
order in chiral perturbation theory.
The corresponding chiral Lagrangian is given by \cite{TV,SS}
\be
{\cal L}_{\rm eff} = \frac 14 F^2 {\rm Tr} \nabla_\nu U
\nabla_\nu U^\dagger - \frac 12 \Sigma{\rm Tr} M(U +U^\dagger),
\ee
with $U \in U(2)$ and 
covariant derivative defined by
\be
\nabla_\nu U = \del_\nu U - i[B_\nu, U]
\ee
with
\be
B_\nu = 
\delta_{\nu,0} 
\mat \mu & 0 \\ 0 & -\mu\emat \qquad {\rm and} \qquad
M = \mat z & 0 \\ 0 & z^* \emat.
\ee
We can neglect higher order terms in $\mu$ as long as $\mu \ll F$.
In the mean field limit, the phase structure of the partition function is
determined by the  static part of the Lagrangian given by
\be
{\cal L }_{\rm stat} = \frac 14 F^2 {\rm Tr} [U,B][U^\dagger, B]
- \frac 12 \Sigma {\rm Tr} M(U +U^\dagger).
\label{leffnf0}
\ee
Notice that $V{\cal L}_{\rm stat}$ only depends on the microscopic variables
(\ref{micro}).
In order for the mean field limit to be valid for the calculation of
bulk quantities,
we also require that
$m, |z| \ll F^4/\Sigma $.

The ground state of this Lagrangian is easily found by making the ansatz
\be
U = \mat \cos \alpha & \sin \alpha \\ -\sin \alpha & \cos \alpha \emat.
\ee
This results in the effective Lagrangian
\be
{\cal L}_{\rm stat} = - 2\mu^2 F^2 \sin^2\alpha - \Sigma(z+z^*)\cos \alpha.
\ee
The minima of this Lagrangian are given by
\be
\sin \alpha = 0 \qquad {\rm and} \qquad \cos\alpha = 
\frac {\Sigma(z+z^*)}{4\mu^2 F^2}
\ee
resulting in the free energies
\be
{\cal L}_{\rm stat}^{\rm N} = -\Sigma(z+z^*)\qquad {\rm and} \qquad {\cal L}_{\rm stat}^{\rm \pi} = -2\mu^2 F^2 
- \frac{\Sigma^2(z+z^*)^2}{8\mu^2 F^2},
\ee
respectively. The boundary between the two phases is given the condition
\be
\Sigma(z+z^*)= 4\mu^2 F^2.
\ee

This condition can easily be understood physically. Since the mass of the 
charged Goldstone bosons is given by
\be
m_\pi^2 = \frac {(z+z^*) \Sigma}{F^2},
\ee
pion condensation occurs when $ \mu^2 > (m_\pi/2)^2 = (z+z^*)\Sigma/(4F^2)$.

A schematic phase diagram of the quenched Dirac spectrum is shown in 
Fig. \ref{fig:rhoq}. 
The normal phase (N), where $\sin\alpha = 0$,
and the pion condensed phase, where $\cos\alpha = 
\Sigma(z+z^*)/(4\mu^2 F^2)$, occur for $|(z+z^*)\Sigma/(4F^2)| > \mu^2$
and $|(z+z^*)\Sigma/(4F^2)| < \mu^2$, respectively. 
Inside the pion condensation region
the eigenvalue density (per unit area) is constant 
\be
\rho_{\pi}^{N_f=0}(z,z^*;\mu)=-\frac 1\pi 
\del_z \del_{z^*}{\cal L}_{\rm stat}^{\rm \pi}
=\frac{\Sigma^2}{4\mu^2 F^2},
\label{rhoQMF}
\ee
while outside this region the eigenvalue 
density is zero $\rho(|x|>2\mu^2F^2/\Sigma)=-\del_z
\del_{z^*}{\cal L}_{\rm stat}^{\rm N} =0$.

\section{Phase Diagram of the Dirac Spectrum for QCD with Dynamical Quarks}
\label{sec:unquench}

In this section we discuss the phase diagram of the Dirac spectrum of 
QCD with dynamical quarks 
at nonzero chemical potential. We will do this by relating our results
to strangeness condensation and pion condensation at nonzero quark chemical 
potentials, a connection that was mentioned in \cite{OSV} and discussed
briefly in the proceedings \cite{osvchina}. 

The discontinuity of  the chiral condensate arises due to the contribution
from the strongly oscillating terms in the spectral density
\cite{OSV,OSVmicro}.  
These terms become exponentially large with increasing volume, 
but are exponentially suppressed elsewhere \cite{O,AOSV}. Here we will show
that this region can be identified from the mean field limit of the chiral
Lagrangian and it therefore shows that our picture of the Dirac spectrum is
valid  beyond the microscopic domain. 

As in the quenched case we use the replica trick to express the unquenched
eigenvalue density  
\be
\rho^{N_f}(z,z^*,m;\mu) &=& \lim_{n\to 0} 
\frac 1V \frac{1}{n} \del_z \del_{z^*} \log 
Z^{n,N_f}(z,z^*,m;\mu),
\ee
in terms of generating functionals for the eigenvalue density
\be
Z^{n,N_f}(z,z^*,m;\mu) &=&\langle {\det}^n(D +\gamma_0\mu +z){\det}^n(D -\gamma_0\mu +z^*){\det}^{N_f}(D +\gamma_0\mu +m)\rangle .
\label{znnf}
\ee 
Below we will see that in the mean field limit of the chiral Lagrangian, the 
free energy does not always scale with $n$ as in the case of the phase quenched
partition function. Therefore, the naive replica trick does not 
necessarily work
in this case. However, it turns out (see section \ref{sec:nrep}) 
that the boundary of the the oscillating region as well as the boundary
of the region with a constant eigenvalues density do not depend on $n$.
Therefore, these boundaries can be obtained from a calculation for $n =1$
which will be discussed in the remainder of this section. 
In the microscopic limit this also follows
from an exact identity (that can be derived from the Toda lattice
equation) between the spectral density and the ratio of the  
the following partition functions (recall that the microscopic variables
defined in (\ref{micro}) are denoted by a hat)  \cite{AOSV}
\be 
\rho^{N_f}(\hat z,\hat z^*,\hat m;\hat \mu) 
& = &  \frac{\hat z \hat z^*}2 
(\hat m^2-\hat z^2)^{N_f}
Z^{n=-1}(\hat z,\hat z^*;\hat \mu)
\frac{Z^{N_f,n=1}(\hat m,\hat z,\hat z^*;\hat \mu)}
{Z^{N_f}(\hat m;\hat \mu)}\ .
\label{rhofinal}
\ee
Here, the label $N_f$ denotes the number of fermionic quark flavors with
mass $m$ and the value of the label $n$ stands for the number of quark
pairs with mass $z$ and $z^*$ (a negative value of $n$ should be interpreted
as $n$ pairs of bosonic quarks). 
Since both the microscopic limit of $Z^{n=-1}(z,z^*;\mu)$ given in
(\ref{zmin1})
 and the $N_f$ flavor partition $Z^{N_f}(\hat m; \hat \mu)
= Z^{N_f}(\hat m)$ do not show a phase transition in the chemical
potential-mass plane, the phase structure of the spectral density 
is determined by 
$Z^{N_f, n=1}(\hat m,\hat z,\hat z^*;\hat \mu)$. Note that the oscillating
nature must come from $Z^{N_f,n=1}(\hat m,\hat z,\hat z^*;\hat \mu)$ since both
$Z^{n=-1}(\hat z,\hat z^*;\hat \mu)$ and $Z^{N_f}(\hat m)$ are real and
positive. The phase structure can be analyzed
in the thermodynamic limit which is smoothly connected to the strong
non-Hermiticity limit ($\hmu\gg1$) of the microscopic domain. In the  
large $\hat \mu$ limit  with $\hat m \sim \hmu^2$
and $|\hat z| \sim \hat \mu^2$  we have that 
\be
 \rho^{N_f}(\hat z,\hat z^*,\hat m;\hat\mu)  
e^{2\hat \mu^2 +\hat x^2/(2\hat \mu^2)+N_f\hat m}
\sim Z^{N_f,n=1}(\hat m, \hat z,\hat z^*;\mu).
\label{rhoz}
\ee
Below we will evaluate the partition function 
$Z^{N_f,n=1}(m,z,z^*;\mu)$ in the mean field limit. The phase structure 
does not depend on the number of
flavors, and  we will only study the case $N_f = 1$. 
Alternatively, one could analyze the strong non-Hermiticity limit of the
spectral density \cite{OSV}
where the phases are determined by a competition between
the quenched and unquenched contribution, and whether a saddle point of
the expression hits the boundary on the integration interval.

\subsection{Low Energy Limit of QCD Partition Function}

The QCD generating functional with $n=1$ for the $N_f=1$ spectral density 
is given by the partition function 
\be
Z^{N_f=1,n=1}(m,z,z^*;\mu) &=& \langle \det (D+\mu \gamma_0+m) \det (D+\mu
\gamma_0+z)  
\det (D^\dagger +\mu \gamma_0+z^*) \rangle \nn \\
&=& \langle \det (D+\mu \gamma_0+m) \det (D+\mu \gamma_0+z) 
\det (D -\mu \gamma_0+z^*) \rangle.
\label{zstrange}
\ee
We can interpret the flavor with mass $m$ as the strange flavor 
and $z$ and
$z^*$ as the up and down flavors. 
The isospin and strangeness chemical
potential are then given by 
\be
\mu_I = 2\mu, \qquad \mu_S = \mu.
\label{chempot}
\ee
The  quark chemical potential $\mu$ of the generating functional 
for the unquenched eigenvalue density 
thus couples to the charges carried by the 
pions and the kaons. That is why the chiral Lagrangian 
can give us nontrivial information
about the Dirac spectrum even though the  Goldstone bosons 
have zero baryon charge. 
The masses of the the Goldstone bosons are equal
to  
\be
m_K^2 = {\rm Re \,}\Sigma(m+z^*)/F^2, \qquad m_\pi^2 = \Sigma(z+z^*)/F^2.
\label{mesonmass}
\ee
For $m \to \infty$ the partition function (\ref{zstrange}) 
reduces to the quenched case discussed in the previous section. In this case
it was found  \cite{misha,TV} that the phase with the
pion  condensate corresponds to the region where the eigenvalue density is nonzero whereas
the region with vanishing  eigenvalue density is in the normal phase
(see eq. ~(\ref{rhoQMF})).  
As we lower the strange quark mass $m$, at some point the mass of the kaon is
less than that of the pion, which leads to the 
formation of a kaon condensate \cite{KT}. Given $m$, we will see below that the
values of $z$ and $z^*$ for which the kaon condensate is nonzero will
correspond to a region in the complex eigenvalue plane where the unquenched
eigenvalue density is oscillating with and amplitude that grows exponentially
with the volume. 

The low-energy limit of the 
partition function (\ref{zstrange}) for real $z$ was analyzed 
in \cite{KT}. Here we extend the arguments to the complex plane. 
The chiral Lagrangian for the Goldstone fields is given 
by the same expression as in the phase quenched case,
\be
{\cal L}_{\rm eff} = \frac 14 F^2 {\rm Tr} \nabla_\nu U
\nabla_\nu U^\dagger - \frac 12 \Sigma{\rm Tr} M(U +U^\dagger),
\label{chiralL}
\ee
with covariant derivative defined by
\be
\nabla_\nu U = \del_\nu U - i[B_\nu, U]
\ee
but  in this case
\be
B_\nu &=& \delta_{\nu,0} {\rm diag}(\frac 13 \mu_B + \frac 12 \mu_I,
\frac 13 \mu_B - \frac 12 \mu_I,\frac 13 \mu_B + \mu_S),\\
{\cal M} &=& {\rm diag}(z,z^*,m),
\ee
and  $U \in SU(3)$.
The phases are determined by the vacuum states that minimize the
free energy. In the mean field limit 
it is sufficient to
consider the zero momentum modes or constant fields, and we only have to
minimize the static part of the effective Lagrangian given by
\be
{\cal L }_{\rm stat} = \frac 14 F^2 {\rm Tr} [U,B][U^\dagger, B]
- \frac 12 \Sigma {\rm Tr} M(U +U^\dagger).
\label{leff}
\ee

\subsection{Vacuum Configurations}

The ground states of (\ref{leff}) can be 
parameterized by  the ansatz
\be
\bar U = R_y(-\beta) R_z(\alpha) R_y(\beta),
\ee
where $R_x(\beta)$ is a rotation by $\beta$ about the $x$-axis, etc.. Because
of (\ref{chempot}) we have $B_0=\mu\,{\rm diag}(1,-1,1)$ 
and the static part of the Lagrangian  becomes
\be
{\cal L}_{\rm stat} =
-2\mu^2F^2(1 -\cos^2\alpha)
 - \Sigma \big [(z-m)\cos^2\beta(\cos\alpha -1)+z +z^*\cos\alpha
 +m\cos\alpha \big] . 
\ee 

 \begin{center}
\begin{figure}[t]
  \unitlength1.0cm
  \begin{center}
  \centerline{\epsfig{file=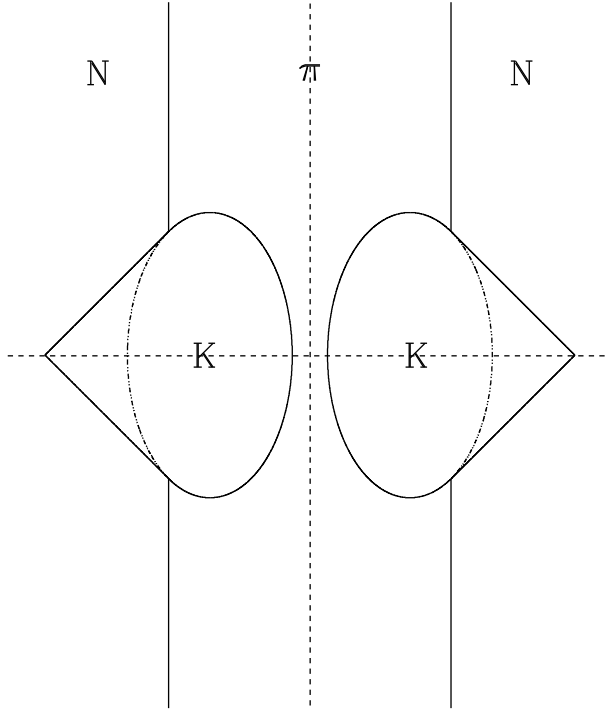,width=8cm}}
  \end{center}
  \caption{  \label{fig:rhoPD}
Phase diagram of the Dirac spectrum in the complex $z$ plane for one dynamical
flavor. The support of
the quenched spectrum is between the two vertical solid black lines. 
Unquenching introduces the oscillations within the ellipses which 
intersect the $x$-axis at $x = \pm m$ and $ x = \pm(\frac 83
F_\pi^2\mu^2/\Sigma -m)$  (dashed curve). The labels N, $\pi$, K refer to 
the normal, pion and kaon condensed phases of the generating functional for 
the eigenvalue density. In the region between the triangle and the ellipse
the spectral density oscillates with an amplitude that is exponentially 
suppressed with the volume.}
\end{figure}
\end{center}

The ground state is given by the minimum of the real and imaginary parts 
of the
Lagrangian. For definiteness we take $m > 0$ and $x \equiv {\rm Re}(z) >0$.
For $\alpha = 0$ the value of $\beta$ is not fixed. This phase, where
$\bar U =1$, is the normal phase with free energy given by
\be
{\cal L}^N_{\rm stat} = -\Sigma(m + z + z^*).
\ee

For $\alpha \ne 0$, the minimum of ${\cal L}_{\rm stat}$ is either at $\beta
= 0$ or at $\beta = \pi /2$. For $\beta=0$ we find 
\be
\cos \alpha = \frac {\Sigma(z+z^*)}{4\mu^2F^2}.
\ee
This is the pion condensed phase with free energy given by
\be
{\cal L}^\pi_{\rm stat}= 
-2\mu^2F^2(1 -\cos^2\alpha)-\Sigma(z+z^*)\cos\alpha - \Sigma m
= -\Sigma m -2\mu^2 F^2 - \frac{\Sigma^2(z+z^*)^2}{8\mu^2F^2}.
\ee
This expression also enters in the relation
between the spectral density and the partition function 
in the strong non-Hermiticity limit  (see Eq. (\ref{rhoz})), and
can be concisely written as
\be
\rho^{N_f=1}_\nu(\hat z,\hat z^*,\hat m;\hmu)
 e^{-V\tilde {\cal L}^\pi_{\rm stat}}
\sim Z^{N_f=1,n=1}(\hat m, \hat z,\hat z^*;\hmu).
\label{rhozL}
\ee
We introduced $\tilde {\cal L}^\pi_{\rm stat}$ as formal expression
for ${\cal L}^\pi_{\rm stat}$ without taking into account
the constraint $|\cos \alpha| \le 1$.
Since $Z^{N_f=1,n=1}(\hat m, \hat z,\hat z^*;\hmu)\sim \exp[-V{\cal L}^\pi_{\rm stat}]$ 
in the pion condensed phase, we observe that the eigenvalue 
density in this region is of
order 1. Alternatively, this relation is necessary 
to obtain  an eigenvalue density that does not depend exponentially
on the volume the in pion condensed
phase.

The mean field result for the eigenvalue density (per unit volume) 
is thus given by
\be
\rho_{\pi}^{N_f=1}(z,z^*,m;\mu)=-\frac 1 \pi\del_z \del_{z^*}
{\cal L}_{\rm stat}^{\rm \pi}
=\frac{\Sigma^2}{4\mu^2 F^2}.
\label{rhoNf1MFpi}   
\ee

For $\beta = \pi/2$ the static free energy is minimized by
\be
\cos \alpha = \frac {\Sigma(m+z^*)}{4\mu^2F^2}\qquad {\rm with}
\qquad {\rm Re}\, {\Sigma(m+z^*)}/({4\mu^2F^2}) < 1.
\label{saddlek}
\ee
This is the kaon condensed phase with free energy given by
\be
{\cal L}^K_{\rm stat}= 
-2\mu^2F^2(1 -\cos^2\alpha)-\Sigma(m+z^*)\cos\alpha - \Sigma z
= -\Sigma z -2\mu^2 F^2 + \frac{\Sigma^2 (m+z^*)^2}{8\mu^2F^2}.\nn\\
\label{LKstat}
\ee
The condition in (\ref{saddlek}) arises from the requirement that
the real part of the saddle point is inside the integration domain.

The phase boundaries occur where the real part of the Lagrangian
for two different phases coincides. Notice that $x > 0$. The boundary
between the normal phase and the pion condensed phase is at
\be
(z+z^*)\Sigma = 4\mu^2 F^2.
\label{normalpi}
\ee
The boundary between the pion condensed phase and the kaon 
condensed phase is given by
\be
m\Sigma  +\frac {(z+z^*)^2\Sigma^2}{8\mu^2 F^2}= x\Sigma + 
{\rm Re} \left[ \frac{(z^*+m)^2\Sigma^2}{8\mu^2F^2}\right ].\ee
With $z=x+i y$
this can be written as
\be 
\frac {x^2\Sigma^2}{2\mu^2F^2} -2x\Sigma +\frac {y^2\Sigma^2}{8\mu^2F^2} 
= -(x+m)\Sigma + \frac{(x+m)^2\Sigma^2}{8\mu^2F^2},
\label{kpion}
\ee
in agreement with \cite{OSV}. The boundary between the normal phase and
the kaon condensed phase is given by
\be
\frac {y^2\Sigma^2}{8\mu^2F^2} = 2\mu^2 F^2-(x+m)\Sigma + 
\frac{(x+m)^2\Sigma^2}{8\mu^2F^2} =
\frac1{8\mu^2 F^2} ((x+m)\Sigma-4\mu^2F^2)^2.
\label{knormal}
\ee
This implies that kaon condensed phase dominates in the region
\be
2\mu^2F^2 < x\Sigma < 4\mu^2F^2 -m\Sigma, \qquad |y\Sigma| 
< 4\mu^2F^2 -(x +m)\Sigma \nn\\
m\Sigma < x\Sigma < 2\mu^2F^2, \qquad \frac {y^2\Sigma^2}{8\mu^2F^2} < 
-(x+m)\Sigma + \frac{(x+m)^2\Sigma^2}{8\mu^2F^2}+2x\Sigma - 
\frac {x^2\Sigma^2}{2\mu^2F^2},
\label{kphase}
\ee
and in a similar region that is given by the mirror image of 
this region with respect to the $y$-axis. Because of the relation
(\ref{rhozL}) the spectral density in the kaon condensed phase is
given by
\be
 \rho^{N_f=1}(\hat z,\hat z^*,\hat m;\hat \mu) \sim
e^{-V{\cal L}^K_{\rm stat}+V \tilde {\cal L}^\pi_{\rm stat}}.
\label{rhozk}
\ee
This takes into account the requirement that
$|\cos  \alpha |\le  1$ in the condensed phase. The expression 
${\rm Re}[{\cal L}^K_{\rm stat}]<{\rm Re}[\tilde {\cal L}^\pi_{\rm  stat}]$
is a curve that splits the kaon condensed region into two parts. Inside 
the ellipse, the eigenvalue density grows exponentially 
with the volume. Since the free energy in the kaon condensed phase is complex, 
cf.~(\ref{LKstat}), the eigenvalue density oscillates with a period
proportional to the inverse volume. Outside of the ellipse defined by the 
second equation in (\ref{kphase}) the oscillations have an exponentially
small amplitude with increasing volume. 

Below we will argue that the boundaries between the different phases
of the Dirac spectrum do not depend on the number of replicas. However,
for $n=1$ the mean field result
for the free energy given by (\ref{LKstat}) results in a vanishing
eigenvalue density in the oscillatory region. This implies that the
resolvent for $n=1$ is discontinuous across the phase boundary between
the pion condensed phase and the kaon condensed phase. We have shown
numerically that the zeros of the $n=1$ partition function are located on
the same ellipse.

The phase diagram in the
$xy$-plane is shown in Fig. \ref{fig:rhoPD}. The kaon condensed region
is split    into two parts by the dotted curve which intersects the 
$x$-axis at $x\Sigma = \frac 83 \mu^2F^2 - \frac 13 m\Sigma$. Only the region
that is inside the ellipse shows exponentially large oscillations.
The three phases intersect each other at
\be
x\Sigma=2\mu^2F^2,\qquad |y|\Sigma = 2\mu^2F^2 -m\Sigma.
\ee

\begin{center}
\begin{figure}[ht]
  \unitlength1.0cm
\centerline{  \epsfig{file=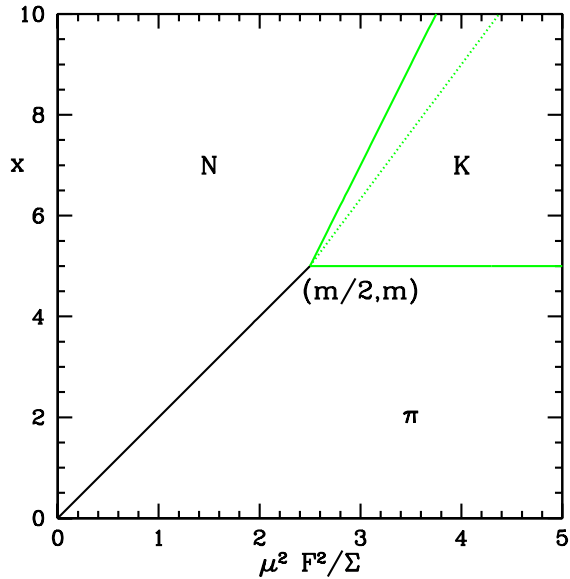,width=8cm}}
  \caption{
Phase diagram of the Dirac spectrum in the $\mu^2-x$ plane.
The dashed curve is the line $x = \frac 83 \mu^2F^2/\Sigma$.}
\label{fig:valanxmu}  
\end{figure}
\end{center}

For $y=0$ we can distinguish the following phases
\cite{KT}: 
\begin{itemize}

\item a normal phase for $2\mu^2F^2 < x\Sigma <m\Sigma$ 
and $2m\Sigma< 4\mu^2F^2 < (x+m)\Sigma$.

\item a pion condensed phase for  $x\Sigma < 2\mu^2 F^2<m\Sigma$ 
and $x \Sigma < m \Sigma$ with $x\Sigma > 2\mu^2 F^2$.

\item a kaon condensed phase for $ 4\mu^2F^2> (x+m)\Sigma > 2m \Sigma $.
\end{itemize}
As we have seen in Fig. \ref{fig:rhoPD}, the kaon condensed 
phase is split into two parts
by the the curve
\be
x\Sigma = \frac 83 \mu^2F^2 - \frac {m\Sigma}3,
\ee
with exponentially large (small) oscillations in the region to the 
right (left) of this curve.
Figure \ref{fig:valanxmu} shows a phase diagram of the 
Dirac spectrum in the $\mu^2-x$ plane for $y=0$ and $\hat m=5$.
For $\mu^2F^2 < m\Sigma /2$ kaon condensed phase is absent.

The border lines between the different phases can be easily understood
physically.
Pion condensation occurs when $\mu_I > m_\pi$ while kaon condensation is
favored when $\mu_S > m_K$ and $m_K < m_\pi$. Using (\ref{chempot})
and (\ref{mesonmass}) the  
condition (\ref{normalpi}) can be rewritten as $\mu_I=m_\pi$. For $y=0$ 
the boundary between the normal phase and the kaon condensed
phase  (see Eq. (\ref{knormal})) can be written as $\mu_S = m_K/2$.
The boundary between the pion condensed phase and the kaon condensed
phase at $y=0$ (see Eq. (\ref{kpion})) can be written as
$m_K= m_\pi$. The reason is that since $\mu_I = 2\mu_S=\mu$, the condensed
phase with the lightest boson will dominate.

\section{Mean Field Free Energy for $n$ Replicas}
\label{sec:nrep}

In this section we derive the mean field free energy of the partition function
(\ref{znnf}). It is obtained from the free energy of the static chiral 
Lagrangian, which is still given by the static part of 
(\ref{chiralL}) but now $U \in U(2n+1)$. The baryon charge matrix and the
mass matrix are now given by
\be
B_\nu &=& \delta_{\nu,0} {\rm diag}({\bf 1_n},-{\bf 1_n},1),\\
{\cal M} &=& {\rm diag}(z {\bf 1_n},z^*{\bf 1_n},m),
\ee
where ${\bf 1_n}$ is the $n\times n$ unit matrix. 

The ansatz for the vacuum configurations is given by 
\be
U = R_{2n+1,p}^{-1}(\beta)
 \left(\begin{array}{ccccccc} 
\cos\alpha_1 & & & \sin\alpha_1 & & & 0\\ 
& \ddots &&& \ddots && \\
&& \cos\alpha_n &&& \sin \alpha_n &\\
-\sin\alpha_1 &&& \cos\alpha_1 &&& \vdots\\
& \ddots &&&\ddots &&\\
&& -\sin\alpha_n &&& \cos\alpha_n& 0\\
0 && & \hdots && 0 & 1\\
\end{array}\right)
R_{2n+1,p}(\beta),
\label{vacn}
\ee
where $R_{2n+1,p}(\beta)$ is a rotation by $\beta$ in the $2n+1,p$-plane
for some choice of $p$.

In the normal phase all $\alpha_k=0$ so that $U=1$ and
\be
{\cal L}^N_{\rm stat} = -\Sigma(m + n z +n z^*).
\ee

For $\alpha_k \ne 0$, the minimum of ${\cal L}_{\rm stat}$ is either at $\beta
= 0$ or at $\beta = \pi /2$. For $\beta=0$ we have
\be
U = 
 \left(\begin{array}{ccccccc} 
\cos\alpha_1 & & & \sin\alpha_1 & & & 0\\ 
& \ddots &&& \ddots && \\
&& \cos\alpha_n &&& \sin \alpha_n &\\
-\sin\alpha_1 &&& \cos\alpha_1 &&& \vdots\\
& \ddots &&&\ddots &&\\
&& -\sin\alpha_n &&& \cos\alpha_n& 0\\
0 && & \hdots && 0 & 1\\
\end{array}\right)
\ee
and find 
\be
{\cal L}^\pi_{\rm stat}= 
-m\Sigma -\sum_{k=1}^n [2\mu^2F^2(1 -\cos^2\alpha_k)+\Sigma(z+z^*)\cos\alpha_k].
\ee
The saddle point equation for each of the replicas is the same with nonzero 
solution given by
\be
\cos \alpha_k = \frac {\Sigma(z+z^*)}{4\mu^2F^2}.
\label{sadpi}
\ee
This is the pion condensed phase with free energy given by
\be
{\cal L}^\pi_{\rm stat}
= -\Sigma m -2n\mu^2 F^2 - n\frac{\Sigma^2(z+z^*)^2}{8\mu^2F^2}.
\ee

For $\beta= \pi/2$, the ansatz (\ref{vacn}) for the vacuum configuration
depends on $p$. To excite
mesons with a nonzero isospin charge we necessarily have that 
$1\le p \le n$. For $k \ne p$ the saddle point equation is again given by
(\ref{sadpi}) and is the same as in the pion condensed case. For $k=p$ the
saddle point equation is given by (\ref{saddlek}) found for $n=1$.
For arbitrary $n$ we thus find that the free energy for the kaon 
condensed phase is given by
\be
{\cal L}^K_{\rm stat}
= -\Sigma z -2n\mu^2 F^2 - (n-1)\frac{\Sigma^2 ( z^*+z)^2}{8\mu^2F^2}
-\frac{\Sigma^2 ( z^*+m)^2}{8\mu^2F^2}.
\ee

\subsection{Phase Boundaries}

It is immediately clear that the phase boundaries between the normal phase
and the pion condensed phase and between the pion condensed phase and 
the kaon condensed phase do not depend on $n$. However, the boundary
between the kaon condensed phase and the normal phase does depend on $n$.
Also for general $n$ the kaon condensed region is split into
a region with a spectral density that grows exponentially with the volume,
and a region where it decreases exponentially with the volume. Because
the spectral density in the pion condensed region does not depend exponentially
on the volume for $n$ replicas, we also have the relation 
\be
\rho^{N_f=1,n}_\nu(\hat z,\hat z^*,\hat m;\hmu)
 e^{-V\tilde {\cal L}^\pi_{\rm stat}}
\sim Z^{N_f=1,n}(\hat m, \hat z,\hat z^*;\hmu).
\label{rhozLn}
\ee
This implies that the boundary of the oscillatory region with an exponentially
large amplitude is given by the equality of the free energies of the 
pion condensed phase and the normal phase and therefore does not depend
on $n$.

\section{Conclusions}

The Dirac spectrum for QCD with dynamical quarks at nonzero quark chemical
potential can be split into three
regions. A region with no eigenvalues, a region with a constant eigenvalue
density equal to the quenched spectral density,
and a region where the spectral density shows oscillations.
This last region is further split into one region with
an exponentially large and one region with an 
exponentially suppressed amplitude.
These regions have been derived from the chiral Lagrangian for the generating
functional of the eigenvalue density. The generating functional includes a
conjugate quark and this makes it possible to reinterpret the chemical
potential as a nonzero isospin and strangeness chemical potential. The three
regions of the eigenvalue density correspond to a normal phase, a pion
condensed phase and a kaon condensed phase of the generating
functional, respectively.  

These results were derived from the leading order chiral Lagrangian
in the mean field limit which is 
valid for $m,|z| \ll F^4/\Sigma $ and $\mu \ll F$. 
However, 
we expect that the Dirac
spectrum beyond this domain will behave in a similar fashion.

The mean field limit is the lowest order term in the standard expansion of
chiral perturbation theory \cite{CPT}. The results presented here thus 
form the basis for an analysis that contains 
loop effects as well as higher order terms in the chiral Lagrangian. The 
inclusion of loop effects in chiral perturbation theory at nonzero chemical
potential is well defined \cite{STV1} and can be extended to include the
effects of a nonzero temperature as well \cite{STV2,DN}.

{\bf Acknowledgments:} 
We wish to thank Poul Henrik Damgaard for useful discussions. This work was
supported in part by U.S. DOE Grant No. DE-FG-88ER40388.

\end{document}